\documentclass[11pt]{article}
\newlength{\vshift}
\newlength{\hshift}
\usepackage{amssymb,amsopn}

\def\la{\lambda}

\def\ka{\kappa}
\def\de{\delta}

\def\si{\sigma}

\def\hp{\hat{\partial}}

\def\h{\hat}
\def\lb{\lbrack}
\def\rb{\rbrack}

\begin{document}

\begin{titlepage}
\rightline{LMU-TPW 2004-06}

\vspace{4em}
\begin{center}

{\Large{\bf Deformed Coordinate Spaces\\
Derivatives}}

\vskip 3em

{{\bf Julius Wess${}^{1,2}$ }}

\vskip 1em
Lecture given at BW2003 Workshop\\
{\it Mathematical, Theoretical and Phenomenological Challenges
Beyond Standard Model}\\
29 August - 02 September, 2003 Vrnja\v cka Banja, Serbia\\

\vskip 1em

${}^{1}$Universit\"at M\"unchen, Fakult\"at f\"ur Physik\\
        Theresienstr.\ 37, D-80333 M\"unchen\\[1em]

${}^{2}$Max-Planck-Institut f\"ur Physik\\
        F\"ohringer Ring 6, D-80805 M\"unchen\\[1em]

\end{center}

\vspace{2em}

\vskip 7.5cm
\qquad\hspace{2mm}\scriptsize{eMail: wess@theorie.physik.uni-muenchen.de}
\vfill

\end{titlepage}\vskip.2cm

\newpage
\setcounter{page}{1}

This lecture  is based on common work with Marija Dimitrijevi\' c,
Larisa Jonke, Frank Meyer, Lutz M\"oller, Efrossini Tsouchnika
and Michael Wohlgenannt \cite{1}.

The aim of this lecture is to clarify the concept of derivatives on
quantum spaces \cite{2}. These derivatives are an essential
input for the construction of deformed field equations
such as the deformed Klein-Gordon or Dirac equations \cite{3}.
These deformed field equations are in turn the starting
point for field theories on quantum spaces.

For a given coordinate space there are in general many
ways to define derivatives \cite{4}. We shall try to develop
a general concept of such derivatives into which all the
different sets of derivatives fit and that allows us by
adding additional requirements - usually based on symmetries -
to reduce the number of possible derivatives.

\vspace{0.5cm}

Let me first remind you of the concept of  {\it deformed coordinate spaces}
(DCS) which we will use
as quantum spaces. DCS are defined in terms of coordinates
$\hat{x}^\mu, \mu=1\dots n$ and relations. Examples of such relations are

1. Canonical relations \cite{5}
\begin{equation}
\lb \h{x}^\mu , \h{x}^\nu\rb = i \theta^{\mu\nu}, \label{1}
\end{equation}
for constant $\theta$ it leads to the socalled
$\theta$-deformed coordinate space ($\theta$-DCS).

2. Lie-type relations \cite{6} where the coordinates form
a Lie algebra
\begin{equation}
\lb \h{x}^\mu , \h{x}^\nu\rb = i C^{\mu\nu}_\lambda\h{x}^\lambda , \label{2}
\end{equation}
$C^{\mu\nu}_\lambda$ are the structure constants. Among these is the
$\kappa$-deformed quantum space ($\kappa$-DCS) \cite{7}.

3. Quantum group relations:
\begin{equation}
\h{x}^\mu\h{x}^\nu =\frac{1}{q}R^{\mu\nu}_{\ \ \rho\sigma}\h{x}^\rho\h{x}^\sigma \label{3}
\end{equation}
where the $R$-matrix defines a quantum group \cite{8}. These
are the $q$-deformed spaces ($q$-DCS).

The DCS is the algebra $\hat{\cal A}_{\hat{x}}$, this is the factor space of
the algebra freely generated by the elements $\hat{x}^\mu$ divided by the
ideal generated by the relations \cite{9}.
We not only consider polynomials in
$\hat{\cal A}_{\hat{x}}$ but formal power series as well.

In short all polynomials of the coordinates $\hat{x}^\mu$ that can be
transformed into each other by using the
relations are linearly dependent.

For the examples listed it can be shown that
the dimensions of the vector spaces of
polynomials with given degree are the same as for
commuting coordinates. This is the socalled
Poincar\' e-Birkhoff-Witt property \cite{10}.

\vspace{0.5cm}

{\it Derivatives} are maps of the DCS
\begin{equation}
\hp : \hat{\cal{A}}_{\hat{x}}\rightarrow \hat{\cal{A}}_{\hat{x}}. \label{4}
\end{equation}
They are usually defined by maps on the coordinates,
and therefore on the free algebra defined by
them. To define a map on the
factor space DCS, derivatives have to be consistent with the relations
defining the DCS. They also should lead to a Leibniz rule.

A very general ansatz for the action of a derivative on the
coordinates is:
\begin{equation}
\lb\hat{\partial}_\mu,\hat{x}^\nu\rb=\de_\mu^\nu + \sum_{j}A^{\nu\rho_1\dots\rho_j}_\mu
\hp _{\rho_1}\dots\hp_{\rho_j}. \label{5}
\end{equation}
The coefficients $A^{\nu\rho_1\dots\rho_j}_\mu$ are complex numbers.
They
have to be chosen such that (\ref{5}) is consistent with the relations.
Having found such coefficients a Leibniz rule can be derived because
$\hp_\mu (\hat{f}\hat{g})$ can be computed from (\ref{5}), $\hat{f}$ and $\hat{g}$ are
elements of $\h{\cal{A}}_{\h{x}}$.

\vspace{0.5cm}

Maps can also be defined on the set of derivatives:
\begin{eqnarray}
E &:& \hp\rightarrow \hp '\nonumber\\
\hp '_\mu &=& E_\mu^{\ \rho}(\hp)\hp _\rho .\label{6}
\end{eqnarray}

The matrix $E$ depends on the derivatives $\hp$ only, not on the
coordinates. Because the derivatives $\hp$ are maps on DCS
the new derivatives will be as well.

If $E$ is invertible and if the matrix $E$ starts with
the Kronecker symbol as derivative-independent term
we obtain from (\ref{6}) again derivatives in the
sense of (\ref{5}). All derivatives satisfying the
consistency condition that have been found
up to now are related by such transformations.

\vspace{0.5cm}

We shall discuss the $\theta$-DCS here, this is the
simplest case. The relations (\ref{1}) are consistent with
\begin{equation}
\lb\hat{\partial}_\mu,\hat{x}^\nu\rb=\de_\mu^\nu .\label{7}
\end{equation}
A short calculation shows:
\begin{equation}
\hp_\rho \big( \hat{x}^\mu\hat{x}^\nu - \hat{x}^\nu\hat{x}^\mu -i\theta^{\mu\nu}\big)
= \big( \hat{x}^\mu\hat{x}^\nu - \hat{x}^\nu\hat{x}^\mu -i\theta^{\mu\nu}\big) \hp_\rho .
\label{8}
\end{equation}
This is sufficient to prove consistency. Equation (\ref{7})
leads to the Leibniz rule by applying (\ref{7}) to the product
of the two functions $\hat{f}\hat{g}$
\begin{equation}
\hp_\mu \big( \hat{f}\hat{g}\big) = (\hp_\mu \hat{f})\hat{g} + \hat{f}(\hp_\mu \hat{g}). \label{9}
\end{equation}

A short calculation shows that
\begin{equation}
\lb \hp_\mu , \hp_\nu\rb = 0 \label{10}
\end{equation}
is  compatible with the relations (\ref{7}). We can assume
that the derivatives commute and define an algebra
that way.

The Leibniz rule (\ref{9}) can be algebraically formulated as
a comultiplication:
\begin{equation}
\Delta \hat{\partial}_\mu = \hat{\partial}_\mu \otimes {\mathbf 1}
 + {\mathbf 1}\otimes \hat{\partial}_\mu . \label{11}
\end{equation}
It is compatible with the Lie algebra (\ref{10}):
\begin{equation}
\lb \Delta\hp_\mu , \Delta\hp_\nu\rb = 0 \label{12}
\end{equation}
and it is coassociative.
Thus, (\ref{10}) and (\ref{11}) define a bialgebra,
the $q$-deformed bialgebra of translations in the
$\theta$-DCS.

Other sets of derivatives can be obtained from
$\hp_\mu$ by a transformation (\ref{6}). In general such
derivatives will not have defining relations
that are linear in $\hp$ such as (\ref{7}). They will also have more
complicated comultiplication rules. Thus the definition
(\ref{7}) singles out a specific type of derivatives. Moreover they will
transform linearly under a $\theta$-deformed orthogonal
or Lorentz group. We shall now show that.

\vspace{0.5cm}

A {\it deformed orthogonal group or a deformed Lorentz group}
will be a deformation of the transformations
\begin{eqnarray}
\lb \delta_\omega, x^\rho\rb &=& -x^\nu\omega_\nu^{\ \rho}, \nonumber \\
\lb \delta_\omega, \partial_\rho\rb &=& \omega_\rho^{\ \nu}\partial_\nu, \label{13}
\end{eqnarray}
where $\omega_\rho^{\ \nu}$ are the parameters of the infinitesimal
orthogonal or Lorentz transformations. The corresponding
Lie algebra satisfies:
\begin{eqnarray}
\lb \delta_\omega, \delta_\omega' \rb &=& \delta_{\omega\times\omega'} \nonumber\\
(\omega\times\omega')_\mu^{\ \nu} &=& -\big( \omega_\mu^{\ \sigma}\omega_\sigma^{'\ \nu} -
\omega_\mu^{'\ \sigma}\omega_\sigma^{\ \nu} \big). \label{14}
\end{eqnarray}
The map (\ref{13}) can be obtained from a differential
operator (angular momentum)
\begin{equation}
\delta_\omega =-x^\nu\omega_\nu^{\ \rho}\partial_\rho .\label{15}
\end{equation}
This concept can be lifted to the $\theta$-DCS.
\begin{equation}
\lb \hat{\delta}_\omega, \hat{x}^\rho\rb = -\hat{x}^\nu\omega_\nu^{\ \rho}
+\frac{i}{2}\big( \theta^{\nu\mu}\omega_\nu^{\ \rho}-\theta^{\nu\rho}\omega_\nu^{\ \mu}\big) \hp_\mu .
\label{16}
\end{equation}
This result was first obtained in \cite{11}. For $\theta=0$ (\ref{16}) agrees with the undeformed
equation (\ref{13}). In
(\ref{16}) coordinates transform into derivatives. The additional
terms are needed to make the deformed Lorentz
transformation compatible with the relation (\ref{1}).
The map $\hat{\delta}_\omega$ is really a map on $\h{\cal{A}}_{\h{x}}$. This can be shown in a short
calculation, applying (\ref{16}) to the relations (\ref{1}). We find
\begin{eqnarray}
&\hat{\delta}_\omega &\hspace*{-0.3cm}\Big( \hat{x}^\mu\hat{x}^\nu -\hat{x}^\nu\hat{x}^\mu -i\theta^{\mu\nu}\Big)
= \Big( \hat{x}^\mu\hat{x}^\nu - \hat{x}^\nu\hat{x}^\mu -i\theta^{\mu\nu}\Big) \hat{\delta}_\omega
\nonumber\\
&&+ \Big( \hat{x}^\rho\hat{x}^\mu -\hat{x}^\mu\hat{x}^\rho -i\theta^{\rho\mu}\Big)\omega_\rho^{\ \nu}
+ \Big( \hat{x}^\nu\hat{x}^\rho - \hat{x}^\rho\hat{x}^\nu -i\theta^{\nu\rho}\Big)\omega_\rho^{\ \mu}.
\label{17}
\end{eqnarray}
Analagous to (\ref{15}) the transformation (\ref{16}) can be generated
by a differential operator
\begin{equation}
\hat{\delta}_\omega =-\hat{x}^\nu\omega_\nu^{\ \rho}\hp_\rho +\frac{i}{2}
\theta^{\rho\mu}\omega_\rho^{\ \nu}\hp_\mu\hp_\nu .\label{18}
\end{equation}
This allows us to calculate the transformations of the
derivatives:
\begin{equation}
\lb \hat{\delta}_\omega, \hp_\rho\rb = \omega_\rho^{\ \mu}\hp_\mu \label{19}
\end{equation}
and the algebraic relations of $\hat{\delta}_\omega$:
\begin{equation}
\lb \hat{\delta}_\omega, \hat{\delta}_\omega' \rb = \hat{\delta}_{\omega\times\omega'}.\label{20}
\end{equation}
That $\hat{\delta}_\omega$ is a map on $\h{\cal{A}}_{\h{x}}$ follows from the fact that $\hp$ and
$\hat{x}$ are. The comultiplication can be calculated by
applying (\ref{18}) to the product of two functions $\hat{f}\hat{g}$. We find:
\begin{equation}
\Delta\hat{\delta}_\omega = \hat{\delta}_\omega\otimes {\mathbf 1} + {\mathbf 1}\otimes\hat{\delta}_\omega
-\frac{i}{2}\big( \theta^{\nu\mu}\omega_\nu^{\ \rho}-\theta^{\nu\rho}\omega_\nu^{\ \mu}\big)
\hp_\rho\otimes\hp_\mu .\label{21}
\end{equation}
This result has recently been obtained by M. Chaichian et al. in \cite{12}.
This coproduct is coassociative because $\hat{\delta}_\omega\hat{f}\hat{g}\hat{h}$ is associative:
\begin{eqnarray}
\hat{\delta}_\omega\hat{f}\hat{g}\hat{h} &=&
(\hat{\delta}_\omega\hat{f}\hat{g})\hat{h}+
\hat{f}\hat{g}(\hat{\delta}_\omega\hat{h})
-\frac{i}{2}\big( \theta^{\nu\mu}\omega_\nu^{\ \rho}-\theta^{\nu\rho}\omega_\nu^{\ \mu}\big)
(\hp_\rho\hat{f}\hat{g})\hp_\mu\hat{h} \nonumber\\
&=&(\hat{\delta}_\omega\hat{f})\hat{g}\hat{h}+
\hat{f}(\hat{\delta}_\omega\hat{g}\hat{h})
-\frac{i}{2}\big( \theta^{\nu\mu}\omega_\nu^{\ \rho}-\theta^{\nu\rho}\omega_\nu^{\ \mu}\big)
(\hp_\rho\hat{f})\hp_\mu(\hat{g}\hat{h}) .\label{22}
\end{eqnarray}

The Lorentz algebra by itself does not form a bialgebra.
Derivatives appear in the comultiplication rule (\ref{21}).
We can, however, interpret (\ref{21}) as a comultiplication
rule for the Poincar\' e algebra (translation included).

Then (\ref{10}), (\ref{19}) and (\ref{20}) define an algebra, the
$\theta$-deformed Poincar\' e algebra with the
comultiplication (\ref{11}) and (\ref{21}). We have obtained
the $\theta$-deformed Poincar\' e bialgebra. The algebra
relations are the same as for the undeformed Poincar\' e algebra,
the comultiplication is deformed.

$\theta$-deformed Poincar\' e bialgebra:
\begin{eqnarray}
\lb \hp_\mu , \hp_\nu\rb &=& 0,\quad \lb \hat{\delta}_\omega, \hp_\rho\rb = \omega_\rho^{\ \mu}\hp_\mu,
\nonumber\\
\lb \hat{\delta}_\omega, \hat{\delta}_\omega' \rb &=& \hat{\delta}_{\omega\times\omega'},\quad
(\omega\times\omega)_\mu^{'\ \nu} = -\big( \omega_\mu^{\ \sigma}\omega_\sigma^{'\ \nu} -
\omega_\mu^{'\ \sigma}\omega_\sigma^{\ \nu} \big), \label{23}\\
\Delta \hat{\partial}_\mu &=& \hat{\partial}_\mu \otimes {\mathbf 1}
 + {\mathbf 1}\otimes \hat{\partial}_\mu, \nonumber\\
\Delta\hat{\delta}_\omega &=& \hat{\delta}_\omega\otimes {\mathbf 1} + {\mathbf 1}\otimes\hat{\delta}_\omega
+\frac{i}{2}\big( \theta^{\mu\nu}\omega_\nu^{\ \rho}-\theta^{\rho\nu}\omega_\nu^{\ \mu}\big)
\hp_\rho\otimes\hp_\mu .\label{24}
\end{eqnarray}
That the algebraic relations and the comultiplication rules
are compatible can be verified directly.

On our way to a field theory we have to define {\it fields}. They
are elements of $\h{\cal{A}}_{\h{x}}$ with certain transformation properties.
For a scalar field $\hat{\phi}$ we define:
\begin{equation}
\hat{\delta}_T\hat{\phi}=-\xi^\mu\hp_\mu\hat{\phi}\> \mbox{ and }\>
\hat{\delta}_L\hat{\phi}=-\hat{\delta}_\omega\hat{\phi}.\label{25}
\end{equation}
The translation is parametrized by the constant vector $\xi^\mu$
and the Lorentz transformation is defined
by the differential operator (\ref{18}). This characterizes $\hat{\phi}$
as a scalar density.

The derivative of a scalar field transforms like:
\begin{equation}
\hat{\delta}_T\hp_\rho\hat{\phi}=\hp_\rho\hat{\delta}_T\hat{\phi}=-\xi^\mu\hp_\mu\hp_\rho\hat{\phi}
\label{26}
\end{equation}
and
\begin{eqnarray}
\hat{\delta}_L\hp_\rho\hat{\phi} &=& \hp_\rho\hat{\delta}_L\hat{\phi} =
-\hat{\delta}_\omega\hp_\rho\hat{\phi}- \lb \hp_\rho,\hat{\delta}_\omega\rb \hat{\phi} \nonumber\\
&=& -\hat{\delta}_\omega(\hp_\rho\hat{\phi})+\omega_\rho^{\ \mu}\hp_\mu\hat{\phi} .\label{27}
\end{eqnarray}
This is the transformation law of a vector field:
\begin{equation}
\hat{\delta}_P\hat{V}_\rho = -\xi^\mu\hp_\mu\hat{V}_\rho -\hat{\delta}_\omega\hat{V}_\rho
+\omega_\rho^{\ \mu}\hat{V}_\mu .\label{28}
\end{equation}

For a tensor or spinor field we define the transformation law
as follows:
\begin{equation}
\hat{\delta}_P\hat{T}_A = -\xi^\mu\hp_\mu\hat{T}_A -\hat{\delta}_\omega\hat{T}_A
+\omega_\rho^{\ \mu}M_{\mu A}^{\ \rho B}\hat{T}_B ,\label{29}
\end{equation}
where $M_{\mu A}^{\ \rho B}$ is a representation of the undeformed Lorentz agebra. It satisfies
\begin{equation}
\lb M^{\rho\si},M^{\ka\la} \rb = \eta^{\rho\la}M^{\si\ka}+\eta^{\si\ka}M^{\rho\la}
-\eta^{\rho\ka}M^{\si\la}-\eta^{\si\la}M^{\rho\ka},
\label{30}
\end{equation}
where $\eta^{\mu\nu}$ is the metric depending on the algebra, Kronecker symbol for $SO(n)$ or Minkowski
metric for $SO(1,n-1)$.
It is easy to see that the transformations (\ref{29}) represent the
algebra (\ref{23}). For the bialgebra we have to specify the comultiplication.
For the translations comultiplication is straightforward:
\begin{eqnarray}
\hat{\delta}_T (\hat{T}_A\otimes\hat{T}_B) &=& -\xi^\mu\Delta(\hp_\mu)(\hat{T}_A\otimes\hat{T}_B)
\nonumber\\
&=&(\hat{\delta}_T \hat{T}_A)\otimes\hat{T}_B) + \hat{T}_A\otimes(\hat{\delta}_T\hat{T}_B). \label{31}
\end{eqnarray}
For the Lorentz transformations we have to use the
comultiplication (\ref{24}). We obtain:
\begin{equation}
\hat{\delta}_L (\hat{T}_A\otimes\hat{T}_B) = (\hat{\delta}_L \hat{T}_A)\otimes\hat{T}_B
+ \hat{T}_A\otimes(\hat{\delta}_L \hat{T}_B)
+\frac{i}{2}\big( \theta^{\mu\nu}\omega_\nu^{\ \rho}-\theta^{\rho\nu}\omega_\nu^{\ \mu}\big)
\hp_\rho\hat{T}_A\otimes\hp_\mu\hat{T}_B .\label{32}
\end{equation}
The compatibility of the algebraic relations
with the comultiplication can again be verified. We
have established a tensor calculus on tensor and spinor
fields.

After these considerations it is clear that the
Klein-Gordon equation and the Dirac equation are covariant

1. Klein-Gordon equation:
\begin{equation}
\big( \eta^{\mu\nu}\hp_\mu\hp_\nu \pm m^2\big)\hat{\phi}=0 .\label{33}
\end{equation}
The sign of $m^2$ depends on the metric $\eta^{\mu\nu}$.

2. Dirac equation:
\begin{equation}
\big(i\gamma^\mu\hp_\mu-m\big)\hat{\psi}=0, \label{34}
\end{equation}
where $\hat{\psi}$ transforms like a spinor and the $\gamma$'s are
the usual $\gamma$ matrices. Invariant Lagrangian with interaction
terms can be constructed with the above tensor
calculus for tensor and spinor fields.

\end{document}